\documentclass[lengthestimate,tighten,amssymb,aps,prl,twocolumn,floatfix,tightenlines,superscriptaddress]{revtex4}

\usepackage{graphicx}
\usepackage{subfigure}
\usepackage{setspace}  
\usepackage{color}
\usepackage{soul}

\newcommand{\ket}[1]{\left | #1 \right \rangle}

\begin{document}
\title{Generating entanglement with low Q-factor microcavities}

\author{A.B.~Young}\email{A.Young@bristol.ac.uk}
\affiliation{Merchant Venturers School of Engineering, Woodland Road
Bristol, BS8 1UB}

\author{C.Y. Hu}\affiliation{Merchant Venturers School of Engineering, Woodland Road Bristol, BS8 1UB}
%
\author{J.G.~Rarity}\affiliation{Merchant Venturers School of Engineering, Woodland Road
Bristol, BS8 1UB}\

\date{\today}

\begin{abstract}
We propose a method of generating entanglement using single photons and electron spins in the regime of resonance scattering. The technique involves matching  the spontaneous emission rate of the spin dipole transition in bulk dielectric to the modified rate of spontaneous emission of the dipole coupled to the fundamental mode of an optical microcavity. We call this regime resonance scattering where interference between the input photons and those scattered by the resonantly coupled dipole transition result in a reflectivity of zero. The contrast between this and the unit reflectivity when the cavity is empty allow us to perform a non demolition measurement of the spin and to non deterministically generate entanglement between photons and spins. The chief advantage of working in the regime of resonance scattering is that the required cavity quality factors are orders of magnitude lower than is required for strong coupling, or Purcell enhancement. This makes engineering a suitable cavity much easier particularly in materials such as diamond where etching high quality factor cavities remains a significant challenge. 
\end{abstract}
\maketitle

Entanglement is a fundamental resource for quantum information tasks, and generating entanglement between different qubit systems such as photons and single electron spins has been shown to be a key to building quantum repeaters, universal gates\cite{PhysRevLett.92.127902, Yao:2004uq,waks:153601,Yao:2005fk, Barrett:2005kx, PhysRevLett.104.160503, PhysRevB.78.085307, PhysRevB.80.205326, PhysRevB.78.125318}, and eventually large scale quantum computers\cite{PhysRevA.78.032318}. These previous proposals for generating entanglement using a deterministic spin photon interface have focussed on having the optical transitions of a spin system strongly coupled to an optical microcavity, or at least deep into the Purcell regime\cite{PhysRevB.78.085307,PhysRevB.78.125318,PhysRevB.80.205326}. Recent measurements in high quality-factor micropillars have suggested that it is hard to fulfil the requirement of strong coupling whilst maintaining the necessary input output coupling efficiency\cite{Young:2011uq}. In order to work around this we propose a non-deterministic spin photon interface that works in the low Q-factor regime where efficient in/out coupling of photons should be possible. The scheme works by operating in a regime of resonance scattering where the decay constants for the optical dipole transitions in bulk dielectric are matched to the decay parameters when resonantly coupled to an optical microcavity.

\begin{figure}
\centering
\includegraphics[width=0.4\textwidth]{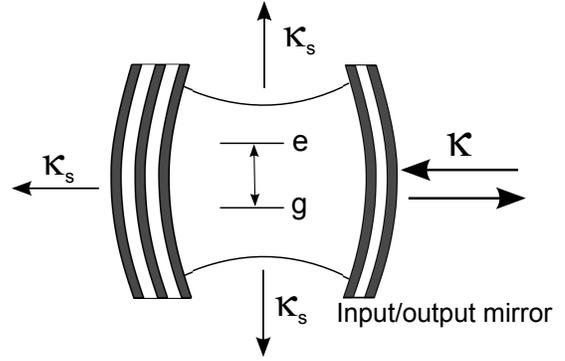}
\caption{Schematic diagram of a single sided cavity coupled to a dipole. $e$ and $g$ represent the excited and ground states of the dipole transition, $\kappa$ represents the coupling rate via the input output mirror and $\kappa_{s}$ represents the loss rate from the cavity system either from the side, transmission through the back mirror, or absorption.}\label{schematic}
\end{figure}

If we consider the single sided dipole-cavity system in Fig.\ref{schematic} then the system can be parameterised by four constants, these are: $\kappa$, the decay rate for intracavity photons via the input/output mirror (outcoupling), $\kappa_{s}$, the decay rate for intra-cavity photons into loss modes, which can include losses out the side of the cavity, transmission and absorption, $g$, the dipole-cavity field coupling rate, and $\gamma$, the linewidth of the dipole transition. We may now express the photon reflectivity when incident on the input/output mirror as\cite{PhysRevB.78.085307}:

\begin{eqnarray}\label{eqn:ref}
&&r(\omega)=|r(\omega)|e^{i\phi}\\
&=&1-\frac{\kappa(i(\omega_{d}-\omega)+\frac{\gamma}{2})}{(i(\omega_{d}-\omega)+\frac{\gamma}{2})(i(\omega_{c}-\omega)+\frac{\kappa}{2}+\frac{\kappa_{s}}{2})+g^{2}}\nonumber
\end{eqnarray}

\noindent where $\omega_{d}$ and $\omega_{c}$ are the frequencies of the QD and cavity, and $\omega$ is the frequency of incident photons. If we match the linewidth of the dipole transition in bulk dielectric ($\gamma$), to the modified spontaneous emission lifetime in the cavity ($4g^{2}/\kappa$)\cite{PhysRevB.60.13276}, then any photons that are input resonant to the dipole-cavity system are scattered into lossy modes. This is due to a destructive interference between the input light and light that is scattered from the dipole.

\begin{figure}
\centering
\includegraphics[width=0.5\textwidth]{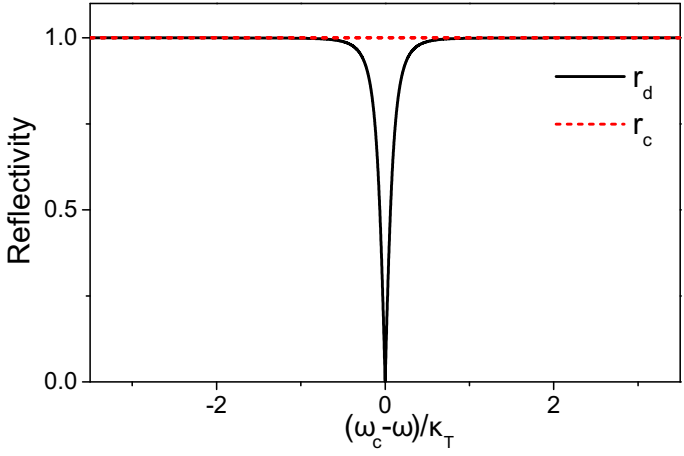}
\caption[Reflectivity of a charged QD cavity in the resonance scattering regime]{Plots showing the reflectivity from an empty cavity ($r_{c}$) and a cavity resonantly coupled to a QD transition ($r_{d}$), using Eqn.\ref{eqn:ref}. The dipole cavity coupling rate has been chosen to be $g^{2}=\gamma\kappa/4$ so that the QD transition resonantly scatters input photons into lossy modes}\label{fig:qdrscat}
\end{figure}

The reflectivity for an empty cavity, and reflectivity for a cavity resonantly coupled to a dipole ($\omega_{d}=\omega_{c}$) can be seen in Fig.\ref{fig:qdrscat}. Here we consider a lossless single sided cavity ($\kappa_{s}=0$), and have set $g^{2}=\gamma\kappa/4$ (the condition for resonance scattering).

We can see that for the case when the dipole transition is resonantly coupled to the cavity then there is a dip in the reflectivity spectrum ($r_{d}$), that goes to zero at zero detuning ($\omega_{c}=\omega_{d}=\omega$). This dip is a result of resonance scattering and has the linewidth of the dipole transition($\gamma$), which we have set to be $\gamma=0.1\kappa$ as an upper limit where $\gamma$ is typically $<<0.1\kappa$ for most atom-cavity \cite{tu-prl-75-4710}, and quantum dot-cavity \cite{nat-432-7014, reitzenstein:251109} experiments. For the case when the cavity is empty ($r_{c}$), all of the input light is reflected. The result is a large intensity contrast between the case of a cavity resonantly coupled to a dipole and an empty cavity. 

If instead of a single dipole transition we coupled a spin system to a cavity in the resonance scattering regime, then if the two dipole transitions corresponding to the $\uparrow$, $\downarrow$ states are distinguishable in some way (energy or polarisation) we can perform a quantum-non-demolition measurement of the spin\cite{Young:2009fk}. From this QND measurement it is possible to generate entanglement non-deterministically between spins and photons. We will now move on to consider some specific spin dipole systems to outline the benefits of generating this non deterministic entanglement in the resonance scattering regime. 

\section{Charged quantum dot in a pillar microcavity.}

We consider the example of a charged quantum dot where the optical transitions for orthogonal spin states couple to orthogonal circular polarisation states of light. By coupling to a pillar microcavity an incident photon would obey the following set of transformations on reflection:

\begin{eqnarray}
\ket{R}\otimes\ket{\uparrow}&\rightarrow& r_{d}\ket{R}\ket{\uparrow}\\
\ket{R}\otimes\ket{\downarrow}&\rightarrow& r_{c}\ket{R}\ket{\downarrow}\\
\ket{L}\otimes\ket{\uparrow}&\rightarrow& r_{c}\ket{L}\ket{\uparrow}\\
\ket{L}\otimes\ket{\downarrow}&\rightarrow& r_{d}\ket{L}\ket{\downarrow}
\end{eqnarray}

\noindent Here if the input photon has right circular polarisation $\ket{R}$, and the spin is in the state $\ket{\uparrow}$ the photon sees a dipole-coupled cavity system and has a reflectivity given by $r_{d}$. Conversely if the spin is in the state $\ket{\downarrow}$ then the input photon sees an empty cavity and has a reflectivity given by $r_{c}$. 

If the input photon has left-circular polarisation $\ket{L}$ then it has the opposite interaction with the spin. In the case when the electron spin of the charged QD is in a equal superposition of spin up and spin down, and two linearly polarised (horizontal) photons are sequentially reflected from the QD-cavity then the output state will be:

\begin{eqnarray}
\nonumber
&&\frac{1}{\sqrt{8}}[(\ket{R}_{1}+\ket{L}_{1})\otimes(\ket{R}_{2}+\ket{L}_{2})\otimes(\ket{\uparrow}+\ket{\downarrow})]\\\nonumber
&&=\frac{1}{\sqrt{8}}(r_{c}^{2}\ket{R}_{1}\ket{R}_{2}+r_{d}^{2}\ket{L}_{1}\ket{L}_{2}+\\\nonumber
&&r_{c}r_{d}\ket{R}_{1}\ket{L}_{2}+r_{d}r_{c}\ket{L}_{1}\ket{R}_{2})\ket{\uparrow}\\\label{eq:pent}
&+&\frac{1}{\sqrt{8}}(r_{c}^{2}\ket{L}_{1}\ket{L}_{2}+r_{d}^{2}\ket{R}_{1}\ket{R}_{2}\\\nonumber
&&+r_{c}r_{d}\ket{R}_{1}\ket{L}_{2}+r_{d}r_{c}\ket{L}_{1}\ket{R}_{2})\ket{\downarrow}
\end{eqnarray}

\noindent Now after a Hadamard pulse ($\pi/2$) on the electron spin we have the state:

\begin{eqnarray}\nonumber
&&\ket{\psi_{out}}=\frac{1}{\sqrt{8}}[(r_{c}^{2}+r_{d}^{2})(\ket{R}_{1}\ket{R}_{2}+\ket{L}_{1}\ket{L}_{2})\\
&&+2r_{c}r_{d}(\ket{R}_{1}\ket{L}_{2}+\ket{L}_{1}\ket{R}_{2})]\ket{\uparrow}\\\nonumber
+&&\frac{1}{\sqrt{8}}[(r_{c}^{2}-r_{d}^{2})\ket{R}_{1}\ket{R}_{2}+(r_{d}^{2}-r_{c}^{2})\ket{L}_{1}\ket{L}_{2}]\ket{\downarrow}
\end{eqnarray}

\noindent From Fig.\ref{fig:qdrscat}, we can see the terms that are proportional to $r_{d}$ will disappear, and $r_{c}=1$. If the electron spin is then measured to be "up" ($\uparrow$) with either a third photon or using the single shot readout technique outlined in previous work\cite{Young:2009fk} then the two photon state will become:

\begin{equation}
\ket{\psi_{out}}=\frac{1}{\sqrt{8}}(\ket{R}_{1}\ket{R}_{2}+\ket{L}_{1}\ket{L}_{2})
\end{equation}

\noindent which is the $\ket{\psi^{+}}$ Bell state. Alternatively if the spin is measured to be down ($\downarrow$) we will project the two photons into the state:

\begin{equation}
\ket{\psi_{out}}=\frac{1}{\sqrt{8}}(\ket{R}_{1}\ket{R}_{2}-\ket{L}_{1}\ket{L}_{2})
\end{equation}

\noindent which is the $\ket{\psi^{-}}$ Bell state. Thus we have generated entangled states with unit fidelity except there is a reduced efficiency of $1/4$. In order to generate larger entangled states then we simply need to reflect more photons from the system however the efficiency scales as $1/2^{n}$, which would make the scheme intractable for entangling large numbers of photons (n). 

\begin{figure}
\centering
\includegraphics[width=0.4\textwidth]{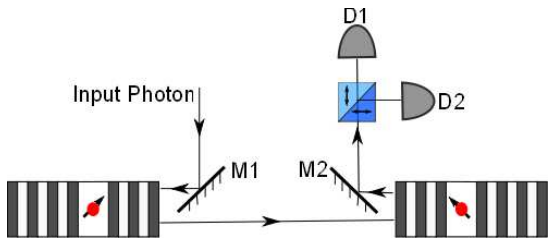}
\caption[Proposed experiment for entangling two charged QD spins]{Schematic diagram of a scheme designed to entangle two spatially separrated spins. There are two charged dots strongly coupled to two spatially separated pillar microcavities. The spins are prepared in an equal superposition state, and a linearly polarised photon is reflected from both. The photon is then split by a polarising beam splitter, upon detecting a H polarised photon (detecter D1), the spins are projected into the $\ket{\Phi^{+}}$ Bell state. If a V polarised photon is detected (detector D2) the spins are projected into the $\ket{\Psi^{-}}$ Bell state.}\label{spinentangler}
\end{figure}

There is an analogous procedure for entangling many spins where photons can be reflected from more than one charged-QD cavity system. Consider the case as in Fig.\ref{spinentangler} where the photon is sequentially reflected from two charged QD-cavity coupled devices operating in the resonance scattering regime. The joint two spin photon state at the output will be 

\begin{eqnarray}
\nonumber
\ket{\psi_{out}}=&&\frac{1}{\sqrt{8}}[(\ket{\uparrow}_{1}+\ket{\downarrow}_{1})\otimes(\ket{\uparrow}_{2}+\ket{\downarrow}_{2})\otimes(\ket{R}+\ket{L})]\\\nonumber
&&=\frac{1}{\sqrt{8}}(r_{c_{1}}r_{c_{2}}\ket{\uparrow}_{1}\ket{\uparrow}_{2}+r_{d_{1}}r_{d_{2}}\ket{\downarrow}_{1}\ket{\downarrow}_{2}\\\nonumber
&&+r_{c_{1}}r_{d_{2}}\ket{\uparrow}_{1}\ket{\downarrow}_{2}+r_{d_{1}}r_{c_{2}}\ket{\downarrow}_{1}\ket{\uparrow}_{2})\ket{R}\\\label{eq:entresscat}
+&&\frac{1}{\sqrt{8}}(r_{c_{1}}r_{c_{2}}\ket{\downarrow}_{1}\ket{\downarrow}_{2}+r_{d_{1}}r_{d_{2}}\ket{\uparrow}_{1}\ket{\uparrow}_{2}\\\nonumber
&&+r_{c_{1}}r_{d_{2}}\ket{\uparrow}_{1}\ket{\downarrow}_{2}+r_{d_{1}}r_{c_{2}}\ket{\downarrow}_{1}\ket{\uparrow}_{2})\ket{L}
\end{eqnarray}

\noindent Where $r_{c_{1}}$, and $r_{c_{2}}$ represent the reflectivity from empty cavity for the first and second cavities respectively, and $r_{d_{1}}$ and $r_{d_{2}}$ represent the reflectivity's from dipole-coupled-cavity systems in the resonant scattering regime for the first and send cavities respectively.

Assuming $r_{c_{1}}=r_{c_{2}}=1$, and $r_{d_{1}}=r_{d_{2}}=0$, if a Hadamard is performed on the photon (i.e. using a polarising beam splitter), upon detection of a horizontally polarised photon, the spins are projected into the state:

\begin{equation}
\ket{\psi_{out}}=\frac{1}{\sqrt{8}}(\ket{\uparrow}_{1}\ket{\uparrow}_{2}+\ket{\downarrow}_{1}\ket{\downarrow}_{2})
\end{equation}

\noindent Which is the $\ket{\psi^{+}}$ Bell state. Alternatively upon detection of a vertically polarised photon, the spins are projected into the state:

\begin{equation}
\ket{\psi_{out}}=\frac{1}{\sqrt{8}}(\ket{\uparrow}_{1}\ket{\uparrow}_{2}-\ket{\downarrow}_{1}\ket{\downarrow}_{2})
\end{equation}

\noindent Which is the $\ket{\psi^{-}}$ Bell state. This is identical to the photonic entanglement generated above and again has an efficiency of $1/4$ associated with photon loss. The benefit of using this technique to entangle spins is that the spin entanglement is heralded upon detection of a photon, thus it is possible to use many photons and keep reflecting them until one is detected. 

\subsection{Entanglement in lossy cavities}

\begin{figure}
\centering
\includegraphics[width=0.5\textwidth]{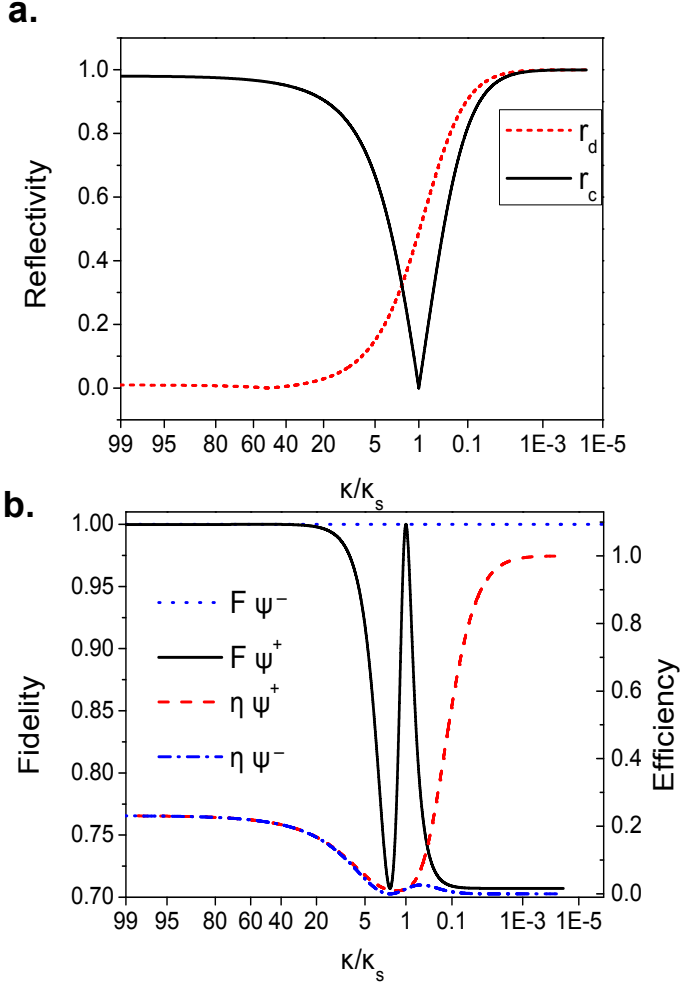}
\caption[Efficiency and fidelity for entanglement generation using resonant scattering]{(a) Showing how the reflectivity at $\omega=\omega_{c}=\omega_{d}$ of an empty cavity ($r_{c}$) and a resonantly coupled charged QD-cavity ($r_{d}$) is affected by changing the ratio of side leakage $\kappa_{s}$ to input-output coupling $\kappa$. The QD-cavity coupling rate has been set to $g^2=\kappa_{T}\gamma/4$ so that we operate in the resonance scattering regime, and the total decay rate $\kappa_{T}=\kappa+\kappa_{s}=const$  (b) Corresponding plot showing how the efficiency $\eta$ and fidelity $F$ of generating the $\ket{\psi^{+}}$ and $\ket{\psi^{-}}$ entangled states is affected by the ratio $\kappa/\kappa_{s}$}\label{resscatF}
\end{figure}

So far to outline this procedure we have assumed that we have a perfect cavity where all the photons escape through the input-output mode ($r_{c}=1$), or are lost through the resonant scattering process, however to make the ideas presented more realistic we must consider cavity imperfections that introduce losses. We must thus include $\kappa_{s}$ in our calculation of the reflectivity. In Fig.\ref{resscatF}.a. we can see a plot of the ratio of the rate of input-output coupling to the rate of losses ($\kappa/\kappa_{s}$) plotted against the reflectivity where the charged QD-cavity is resonantly coupled, and the probe photons are resonant with both ($\omega_{d}=\omega_{c}=\omega$). In this plot the Q-factor of the cavity remains constant, i.e. the total decay rate is not changed ($(\kappa+\kappa_{s})/\kappa_{T}=1$). We have set $g=\sqrt{\kappa_{T}\gamma/4}$, and set $\gamma=0.1\kappa$

\noindent Let us first consider the case of an empty cavity given by the line $r_{c}$ in Fig.\ref{resscatF}.a. (black line) Here we can see that in the regime where $\kappa>>\kappa_{s}$, the reflectivity at zero detuning ($\omega_{c}=\omega$) is $\approx1$. As $\kappa_{s}$ is increased then the reflectivity on resonance drops corresponding to more light being lost from the cavity, until the point when $\kappa_{s}=\kappa$ at which point the reflectivity on resonance drops to $r_{c}=0$, this corresponds to the cavity resonantly transmitting light into lossy modes. As $\kappa_{s}$ is increased further then the coupling into the cavity becomes poorer, until in the regime when $\kappa_{s}>>\kappa$, when the coupling via the input output mode is negligible and the cavity behaves as a conventional mirror.  For a charged quantum dot where the dipole transitions are resonantly coupled to a cavity (red dashed line), in the regime that $\kappa>>\kappa_{s}$ then  $r_{d}=0$. This is what we expect to observe for the case of a resonantly coupled charged QD-cavity in the resonance scattering regime where input photons destructively interfere with scattered photons, and all of the light is lost to non-cavity modes. As $\kappa_{s}$ is increased an extra damping term is added the result is the destructive interference is no longer perfect and some light is reflected. As $\kappa_{s}$ is increased further it begins to dominate and the interference becomes constructive and the reflectivity from a dipole coupled cavity system ($r_{d}$), becomes greater than that of an empty cavity ($r_{c}$). In the limit when $\kappa_{s}>>\kappa$ then no light enters the cavity thus no light is scattered by the dipole transition and we have that $r_{d}=r_{c}$.

The effect of losses on the fidelity is that the terms proportional to $r_{d}$ in Eqn.\ref{eq:entresscat} no longer disappear and the $\ket{\psi^{+}}$ entangled state is no longer prepared with unit fidelity, but instead with a reduced fidelity given by:

\begin{equation}
F_{\psi^{+}}=\frac{1}{\sqrt{1+\frac{4(r_{d}r_{c})^{2}}{(r_{d}^{2}+r_{c}^{2})^{2}}}}
\end{equation}

\noindent for the case when we wish to entangle two photons with one spin with an efficiency $\eta_{\psi^{+}}$ given by:

\begin{equation}\label{eq:resscateff}
\eta_{\psi^{+}}=\frac{(r_{d}^{2}+r_{c}^{2})^{2}}{4}
\end{equation}

\noindent For the case when we wish to entangle two spins with one photon then we have to slightly modify these equations so that fidelity is now:

\begin{equation}
F_{\psi^{+}}=\frac{1}{\sqrt{1+\frac{2(r_{d_{1}}r_{c_{2}})^{2}+2(r_{c_{1}}r_{d_{2}})^{2}}{(r_{d_{1}}r_{d_{2}}+r_{c_{1}}r_{c_{2}})^{2}}}}
\end{equation}

\noindent where the efficiency is now:

\begin{equation}\label{eq:resscateff2}
\eta_{\psi^{+}}=\frac{(r_{d_{1}}r_{d_{2}}+r_{c_{1}}r_{c_{2}})^{2}}{4}
\end{equation}

\noindent Note that the fidelity is not influenced by the two charged-QD cavity systems having non equal values of $r_{c}$ and $r_{d}$, but only by the intensity contrast at both individual dipole cavity system. This means that both systems need not be identical a great advantage when it comes to fabrication of such structures. 

The preparation of the $\ket{\psi^{-}}$ state is not affected by changes in $r_{d}$, and $r_{c}$ and always has $F=1$, but has an efficiency given by:

\begin{equation}
\eta_{\psi^{-}}=\frac{(r_{d}^{2}-r_{c}^{2})^{2}}{4}
\end{equation}

In Fig.\ref{resscatF}.b we can see a corresponding plot for how the fidelity and efficiency is affected by changing the ratio of $\kappa/\kappa_{s}$. Note we have maintained an overall $\kappa_{T}=const$, thus the Q-factor is constant. At the point where $r_{d}=r_{c}$ ($\kappa\approx 2\kappa_{s}$) there is a minimum in fidelity for the preparation of the $\ket{\psi^{+}}$ state, as at this point the cross terms proportional to $\ket{R}_{1}\ket{L}_{2}+\ket{L}_{1}\ket{R}_{2}$ in Eqn.\ref{eq:pent} are a maximum. When $\kappa=\kappa_{s}$ the reflectivity for an empty cavity is zero ($r_{c}=0$), therefore there is a peak in the fidelity and $F=1$, however the since $r_{d}\approx0.5$ the efficiency is low ($\eta\approx0.016$). As we move into the region where $\kappa<\kappa_{s}$ then both $r_{d}$ and $r_{c}$ increase and the efficiency increases, $r_{c}$ increases faster than $r_{d}$, until the limit when $\kappa<<\kappa_{s}$ where $r_{d}=r_{c}=1$ and $\eta=1$. However in this regime there is a minimum in fidelity ($F=1/\sqrt{2}$) for the preparation of the $\ket{\psi^{+}}$ state, again due to the two reflectivities being equal. Note the fidelity for the preparation of the $\ket{\psi^{-}}$ state remains $F=1$, however in a cavity with a large ratio of leaks to input-output coupling the efficiency $\eta_{\psi^{-}}$ drops to zero.  

In order to achieve entanglement with the highest possible efficiency and fidelity for both $\ket{\psi^{+}}$ and $\ket{\psi^{-}}$ states it is necessary to have $\kappa>>\kappa_{s}$. This requirement at first sight is no different to the requirement for the deterministic spin photon-interface outlined in previous work\cite{PhysRevB.78.085307,PhysRevB.78.125318,PhysRevB.80.205326,waks:153601}. So seemingly the non-deterministic scheme outlined offers no advantage, however the required cavity Q-factors are significantly less. In order to see some of the benefits of entanglement generation using resonance scattering it is necessary to consider in more detail some experimental parameters.

We consider some of the state of the art QD pillar microcavity experiments performed by Reithmaier et. al (2004)\cite{nat-432-7014}. Here they showed strong coupling of a QD to a pillar microcavity where the QD-cavity coupling rate $g=80\mu$eV, the cavity linewidth was $\kappa_{T}=180\mu$ev (Q=7350), and the QD linewidth was $\gamma<10\mu$eV at low temperature. If we now assume the maximum value for the QD linewidth ($\gamma=10\mu$eV), then the required cavity linewidth in order to fulfil the requirement for resonance scattering is $\kappa_{T}=2.56$meV (Q=517). This a significantly smaller value than would be required for a deterministic spin-photon interface in previous work\cite{PhysRevLett.92.127902, PhysRevB.78.085307} where we would require $g>\kappa_{T}+\gamma$, meaning $\kappa_{T}=70\mu$ev (Q=18900). With the reduced Q-factor that is required to generate entanglement with resonance scattering, comes a secondary crucial benefit. The state of the art micropillars used in the experiment above and most high-Q micropillars, are limited by losses. Small diameter high Q micropillars have significant sidewall scattering and operate in the regime where $\kappa<\kappa_{s}$. Assuming the linewidth of the pillar is entirely defined by losses out the side $\kappa_{T}=\kappa_{s}=180\mu$eV. The Q-factor can then be reduced by removing, or growing fewer DBR mirror pairs. This will increase $\kappa$ whilst $\kappa_{s}$ should remain constant. Reducing the Q-factor in such a way so that $Q=517$, would result in $\kappa=2.38$meV, and $\kappa_{s}=180\mu$eV, thus have $\kappa/\kappa_{s}=13$. So by reducing the Q-factor we simultaneously increase the input-output coupling rate and move into a regime where the losses out of the side of the pillar become negligible. This means that we can entangle two spins or two photons using charged QD's coupled to such cavities, with fidelity $F>99\%$ in both the $\ket{\psi^{+}}$ and $\ket{\psi^{-}}$ states, with an efficiency $\eta=0.14$.

We have already discussed that this scheme is best employed when used to herald entanglement between many spins. Assuming perfect detection it would be necessary to send in $\approx10$ photons to ensure one was detected heralding the entanglement of two spins. The photons would have to be separated by a time greater than the spontaneous emission lifetime of the QD $\approx1ns$, so it would take approximately $10ns$ to entangle two spins. Pairs of spins could be entangled in parallel and then entanglement could be generated between pairs by repeating the process between single spins from each pair. Hence a linear cluster of N spins could be entangled in $\approx 20ns$, well within the $\mu$s coherence time of a charged QD spin. By parallelising the entanglement procedure we compensate for the non-deterministic nature of generating entanglement using resonance scattering at the expense of the complexity of the photon source required to perform the experiment. 

The advantage of the non-deterministic scheme for generating entanglement is that clearly the required Q-factor is low. A knock on effect is that low Q-micropillars naturally have good input-output coupling efficiency and it is easy to achieve $\kappa>>\kappa_{s}$. To realise the spin-photon interface in the strong coupling regime requires high Q-factor low loss pillars which are much more challenging to fabricate. Further the low Q-factor means the spectral width of the cavity is large compared to the linewidth of the dipole transitions $\gamma$. This means that charged QD's in different micropillar samples have a larger range over which they can be tuned and still be resonantly coupled to the microcavity meaning it will be easier to realise the situation where both dipole transitions are at the same wavelength. Finally the low Q-factor will lessen the effects of any ellipticity or mode splitting in the cavity. Since the linewidth of the $E_{x}$ and $E_{y}$ modes will be large then any mode-spilitting as a result of fabrication error would be small in comparison

The downside to operating in the regime of resonance scattering is that the charged QD-pillar system has to be engineered so that $g^2=\kappa_{T}\gamma/4$. Since the position of self-assembled QD's is random, fulfilling this requirement will be difficult, and may require the growth of site controlled QD's with pillars etched out of the wafer around them. This is not a problem for the spin-photon interface in the strong coupling regime where the coupling rate $g$ just has to be above the threshold where $g>\kappa_{T}+\gamma$, but not have a specific value. Hence operating in the resonance scattering regime changes the nature of the engineering problem. It is easy to achieve a low loss micropillar, but it will be difficult to precisely control the structure to meet the condition for resonance scattering. One possible system that would lend itself to this sort of technique could be toroidal,or microsphere cavities where the Q-factor can effectively be tuned by changing the distance between the cavity and an evanescently coupled tapered fiber. It remains to be seen if the realisation of the structures required for this non-deterministic entanglement scheme will be any easier than the structures required for deterministic spin photon interface in the strong coupling regime. 

\section{application to NV center in diamond}

\begin{figure}
\centering
\includegraphics[width=0.5\textwidth]{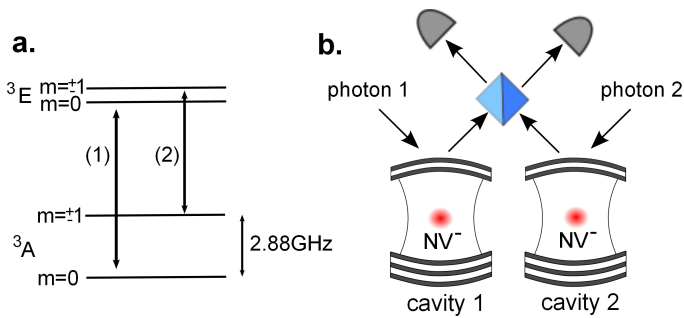}
\caption{{\bf a.} Energy Level diagram for $NV^{-}$ colour center in diamond showing the ground state is splitting\cite{fedor,loubser}. {\bf b.} Schematic diagram of a Barrett and Kok\cite{Barrett:2005kx} style scheme to entangle two NV centres coupled to optical microcavities in the resonance scattering regime. Photons 1 and 2 have energy corresponding to transition 1 ($\hbar\omega_{m=0}$) and are reflected from cavities 1 and 2 respectively and then interfered on a 50:50 beamsplitter.}\label{nvscheme}
\end{figure}

The entanglement protocol outlined here for charged QD-spins, could be applied to other spin systems for example the NV-center in a photonic crystal\cite{Young:2009fk}. Here distinguishing between the two spin states can be achieved with frequency instead of polarisation. If photons were passed through an electro-optical modulator then they can be placed in a superposition of two distinct frequencies $A$ and $B$. Frequency $A$ can then be tuned to be resonant with the $^{3}A_{(m=0)} \rightarrow ^{3}E$ transition (transition (1) Fig.\ref{nvscheme}.a.), and frequency $B$ resonant with the $^{3}A_{(m=\pm1)} \rightarrow^{3}E$ transition (transition (2) Fig.\ref{nvscheme}.a). Since the linewidth of the zero phonon line at low temperature is of order MHz\cite{0953-8984-18-21-S08} then there will be two dips in the reflectivity as a result of resonance scattering corresponding to the $m=0$ and $m=\pm1$ spin states of order MHz spilt by $\approx 2.88$GHz. The distinguishability of these two dips allows us to perform a quantum non-demolition measurement of the spin\cite{Young:2009fk}, and generate entanglement using precisely the same protocol as outlined for the case of a charged quantum dot using photons in a superposition of frequency instead of polarisation. 

Recent results\cite{Togan:2010fk} have also shown that the $m=\pm1$ spin states can be used as a qubit and orthogonal circular polarisations of light then couple the ground states to an excited state $A_{2}$. In this instance the resonant scattering protocol outlined for the charged QD could be directly applied to a NV-center coupled to an appropriate optical microcavity.

An alternative method to generating entanglement in this regime that is perhaps simpler for the case of the NV-center is to only use photons with frequency $\omega_{m=0}$ that are resonant with transition 1 in Fig.\ref{nvscheme}.a. In Fig.\ref{nvscheme}.b. we can see a schematic diagram of how this could work. We can take two photons 1 and 2 that are both resonant with the spin preserving transition 1, and reflect them from two cavity systems 1 and 2 that are both coupled to an NV-center in the resonance scattering regime. After the two photons are reflected they are then interfered on a 50:50 beamsplitter. The entanglement would then be generated using the exact same protocol as outlined by Barrett and Kok\cite{Barrett:2005kx}, which could lead to the formation of large cluster states. One benefit of realising this type of scheme using a resonance scattering technique is that we do not need to use photons that are generated via spontaneous emission from spin in the cavity, and can use some external source, in fact photon 1 and photon 2 can be produced from the same source. This means it should be easier to ensure that the two photons are indistinguishable, which remains a challenge\cite{Bernien:2012fk}, thus effectively removing a decoherence channel from the existing Barrett and Kok protocol. Further to produce indistinguishable phonons via spontaneous emission would require the photons produced to be transform limited. This would require some Purcell enhancement thus $g^{2}>\kappa_{T}\gamma/4$ hence the Q-factor required would need to be higher. Note that this technique is also possible for other spin cavity systems for example the charged QD system examined earlier where we would just set photons 1 and 2 to have the same circular polarisation. 

Finally for illustrative purposes we can consider coupling a nitrogen vacancy centre to a photonic crystal cavity with current state of the art fabrication techniques. Recent results have shown\cite{Riedrich-Moller:2012kx} the fabrication of photonic crystals in diamond with Q-factor of $\approx700$ and a mode volume of $\approx 0.13\mu$m$^{3}$. Using this mode volume and given a typical oscillator strength for the ground to excited state triplet transitions of $f\approx0.12$\cite{nat_phs_2_408}, then we can calculate the dipole-cavity coupling rate to be $g\approx13.5\mu$eV. Since the zero phonon linewidth at low temperature is $\gamma\approx0.1\mu$eV\cite{0953-8984-16-30-R03} then the Q-factor required to meet the resonance scattering condition in such a structure would be $Q\approx256$ nearly three times smaller than has already been experimentally realised. So provided the input/output coupling rate $\kappa$ can be made much larger than the loss rate $\kappa_{s}$ then current experimentally realised structures in diamond would be suitable for generating entanglement using resonant scattering techniques.


\section{Summary}

In Summary we have shown a way to non-deterministically generate entanglement between electron spins and photons. We have shown how this can be applied to charged QD-spins, and nitrogen vacancy centers coupled to optical microcavities. The idea uses resonance scattering where orthogonal photon states are scattered and lost depending on the internal spin state of the electron spin. The advantage to this scheme is that it requires low Q micropillars where the input-output coupling rate is intrinsically high. The disadvantage is the non-deterministic nature makes scaling difficult compared to the spin-photon interface in the strong coupling regime. 


\bibliographystyle{unsrtnat}

\end{document}